\documentclass{aa}
\usepackage{graphicx}

\begin{document}

\title{On the broadening of emission lines in active galactic nuclei}

\author{Luigi Foschini}
\institute{Istituto di Astrofisica Spaziale e Fisica Cosmica
(IASF--CNR), Sezione di Bologna\thanks{Formerly Institute TeSRE -- CNR}, Via
Gobetti 101, I--40129, Bologna (Italy)}
\offprints{\texttt{foschini@tesre.bo.cnr.it}}
\date{Received 10 July 2001; Accepted 4 February 2002}

\abstract{Recent works, both theoretical and observational, have
suggested that turbulence could play a non--negligible role in the broadening
of emission lines in active galactic nuclei. The purpose of this note is to
show how shock wave--turbulence interaction, under unsteady regime, can affect
the broadening of emission lines.
}
\maketitle

\section{Introduction}
The Seyfert galaxies are a class of active galactic nuclei (AGN) characterized
by nuclear spectra with strong emission lines by highly ionized atoms. These
lines are of two types: narrow, corresponding to low density plasma with bulk
velocities of the order of 100 km/s; broad, corresponding to high density
plasma and bulk velocities of the order of $10^4$ km/s. The main physical
mechanism responsible for the ionization is the non--stellar photon flux
coming from the AGN itself (for reviews, see Davidson \& Netzer 1979, Krolik
1999, Fabian et al. 2000, Sulentic et al. 2000). However, also shocks can
provide the physical conditions necessary to the ionization of the atoms. This
occurs, for example, when the postshock temperature reaches values higher than
1 eV, and therefore the shocked gas emits a continuum in the EUV--soft X--ray
band, which is quickly absorbed by nearby gas and its energy converted into
emission lines. See also, however, the work of Perry \& Dyson (1985),
Collin--Souffrin et al. (1988), the former recently
updated and enlarged by Fromerth \& Melia (2001).

The main objection to shock formation of line emission is that the wide range
of observed ionization stages is not fully compatible with a thermal emission
from post--shock regions. The temperature required to produce the highest
stages are far from those obtained in shocks. In addition, according to
Krolik (1999), we have no direct evidence of the existence of shocks, but
other authors suggested that shocks can be important in the morphology,
kinematics, and excitation of emission lines in Seyfert galaxies (Colbert et
al. 1998, Allen et al. 1999). Indeed, the emission line ratios do not
demonstrate the excitation mechanism (shocks or otherwise).
Moreover, the velocity field and line width are consistent with shocks and
with energetics determined from radio and X--ray observations (Allen et al.
1999).

Turbulence has been invoked to explain the line broadening (Horne 1995, Pariev
\& Bromley 1998), but some authors object that the turbulent velocity field
in an AGN disk is not so strong to affect the line profile (Fabian et al.
2000). Bottorff et al. (2000) have found that extra thermal,
non--dissipative microturbulence somewhere between $0$ and $10^3$ km/s, is
consistent with observations of a small sample of quasars with high signal to
noise spectra (Baldwin et al. 1996). In addition, Bottorff \& Ferland (private communication; to
be published on ApJ, April 1, 2002 issue) find that $\approx 200$ km/s microturbulence
with dissipation is able to explain typical emission line ratios found in quasars
and other AGN. Observations poorly constrain the range of turbulence in AGN, but
even the possibility of its presence has important ramifications in the analysis of
AGN emission lines. It is therefore useful to investigate mechanisms that generate
and sustain (possibly high levels of) turbulence.

The purpose of this note is to discuss how the interplay between shock waves
and turbulence can enhance the turbulent velocity field. These physical issues
have been already applied to the line broadening in pulsating stars
(Gillet et al. 1998) and we try to apply these concepts to active galactic
nuclei.

\section{Fully developed turbulence}
Before going on, let us to remember some concepts in fully developed
turbulence. We refer to Landau \& Lifshitz (1987). Turbulent flow is
characterized by random variations in space and time in the velocity field
and develops at very high Reynolds number:

\begin{equation}
Re=\frac{lV\rho}{\mu}
\label{e:rey}
\end{equation}

\noindent where $l$ is the characteristic dimension of the flow, $V$ is the
velocity, $\rho$ the mass density, and $\mu$ the viscosity, which is equal to
about $10^{-17}T^{5/2}$ $\mathrm{g}\cdot \mathrm{cm}^{-1} \mathrm{s}^{-1}$ for
a fully ionized unmagnetized hydrogen plasma (see Choudhuri 1998). For a
typical electron temperature of $10^4$ K, we have $\mu = 10^{-7}$
$\mathrm{g}\cdot \mathrm{cm}^{-1} \mathrm{s}^{-1}$.

By considering a typical flow speed of $V_2$ km/s, where $V_2=V/100$ km/s,
and a mean density of $1.67\cdot 10^{-15}$ g/cm$^3$ (i.e. an electron volume density of $10^9$
cm$^{-3}$ for a fully ionized hydrogen plasma), we obtain from Eq.
(\ref{e:rey}) that $Re\approx 0.167\cdot l \cdot V_2$, where $l$ is in cm. Therefore,
when the characteristic dimension of the flow is of some hundreds of metres we have
already a turbulent flow (for a speed of the order of 100 km/s). On astrophysical
scales, we can consider that the characteristic length is of the order of
astronomical units (AU), so that the flow is fully turbulent, even though in
presence of tiny values of viscosity.

In a turbulent regime, we can divide the velocity $\vec{V}$ into a mean value
component $\bar{\vec{v}}$, obtained from an average over a long time, plus a
fluctuating part $\vec{u}$ that when averaged is equal to zero:

\begin{equation}
\vec{V}=\bar{\vec{v}}+\vec{u}
\label{e:turbo1}
\end{equation}

The order of magnitude of the distance over which the speed varies
appreciably (i.e., the dimension of the largest eddies) has the same order of
the dimension $l$ of the region where there is the flow. Therefore, the
fluctuation of the speed is of the same order of the variation of the mean
speed over the scale $l$. However, under normal conditions, the speed
fluctuations cannot exceed the local sound speed $c_s$ (cf. Balbus \& Hawley
1998). With respect to time, the velocity does not vary appreciably over a
timescale smaller than about $l/\bar{v}$.

Since the viscosity decreases as the Reynolds number increases, in larger
eddies there is no or negligible viscosity, and hence no energy dissipation.
The energy flows from larger to smaller eddies, where it is transformed into
heat and dissipated. To maintain a steady state flow, an external source of
energy must be provided to supply energy to large eddies.

In the plasma of an accretion disk around a supermassive black hole, in
addition to the fluid viscosity, other effects take place, because of the
nature of the plasma. In this case, the electromagnetic forces play an
important role in transferring energy and momentum. The change in state,
particularly during the passage of a shock, derives from the collective
interactions between charged particles and electromagnetic fields. The fields
can be mainly of two types: constant in time (i.e. produced by space charge
separation, currents, or other external sources) or fluctuating in time (i.e.
produced by plasma instabilities). The first case is usually referred as
laminar, while the second as turbulent. For more details see, for example,
Tidman \& Krall (1971).

\section{Shock waves--turbulence interaction}
Despite its importance in several application of science and engineering, the
interaction of shock waves with turbulence is still largely unknown. The first
problem is to obtain reliable experimental data at high Mach numbers ($M > 5$,
hypersonic flow), but also experiments at moderate Mach numbers suffer with a
strong dependence on the measurement instruments and therefore cannot describe
with reasonable reliability the complex interaction of shock waves with
turbulence.

Recently, the availability of computer power made it easier to set
up numerical models, but problems remain, because, when dealing with
turbulence, it is necessary to introduce closure conditions, which in turn are
dependent on experimental data. For a review, see Andreopulos et al.~(2000),
Lele~(1994), Adamson \& Messiter~(1980), and references therein.

However, there is a general consensus that the most important feature of
the interaction between shock waves and turbulence under unsteady
conditions is the amplification of speed fluctuations and a strong change in
the scale lengths, even though it is not well understood how
this amplification occurs (Andreopulos et al.~2000). According to actual
studies, the amplification depends on the shock strength, the state of the
turbulence, and its level of compressibility. With respect to the
amplification of the turbulent kinetic energy, Rotman~(1991) reports an
increase of a factor $2-2.15$, while for Jacquin et al.~(1993) the
amplification depends on the density ratio:

\begin{equation}
\frac{k^2(t)}{k^2(0)} = \frac{2+\rho^2(t)/\rho^2(0)}{3}
\label{e:ampli}
\end{equation}

\noindent where $k(t)$ and $\rho(t)$ are the turbulent kinetic energy and the
gas density at the time $t$, respectively. In this case, the amplification
of the kinetic energy can be up to $\sqrt 6$ for monatomic gases (see the next
Section).

It is worth remembering, that at high Mach number and for steady state
motion, the effect of compressibility suppresses the turbulence and therefore
there is no amplification. Changes in the Mach number (unsteady motion)
generate distortions in shock waves, allowing the interaction with turbulence.
\emph{The key point is therefore that the Mach number must not be constant}.
According to Jacquin et al.~(1993), these conditions represent the ``pressure
released'' regime and are the \emph{upper limit} of the amplification. In this
case, the turbulent Mach number can be higher than 1.

\section{Post--shock conditions}
For the purposes of the present note, we rearrange the Rankine--Hugoniot
relations in the limit of a strong shock, that is with Mach number $M>>1$ (for
reviews see, for example, Zel'dovich \& Raizer 1967, Landau \& Lifshitz 1987,
Frank et al. 1992, Krolik 1999). The post--shock pressure $P_2$ is:

\begin{equation}
P_{2}\approx (1-\frac{\rho_1}{\rho_2}) (1+\alpha)\rho_{1} V_{1}^{2}
\label{e:state1}
\end{equation}

\noindent where $\alpha$ is the ionization factor ($\alpha=1$ full ionization),
$\rho_{2}$ is the post--shock density, $\rho_{1}$ and $V_{1}$ are the
undisturbed density and speed of the plasma. Here, and in the following we use the
subscripts $1$ and $2$ to indicate the undisturbed and the post--shock
values, respectively.

The density ratio across the shock is given once only the value of the
specific heat ratio $\gamma$ is known:

\begin{equation}
\frac{\rho_2}{\rho_1} \approx \frac{\gamma+1}{\gamma-1}
\label{e:shock1}
\end{equation}

From the equation of continuity across a shock, we can get easily that the ratio
of velocities is a function of $\gamma$ only:

\begin{equation}
\frac{V_1}{V_2} = \frac{\rho_2}{\rho_1} \approx \frac{\gamma+1}{\gamma-1}
\label{e:shock2}
\end{equation}

In studies of AGN, velocities widths of typical bright BLR emission lines
are non relativistic, therefore $\gamma$ is generally considered to be $5/3$,
the value of monatomic gas or metal vapors. Therefore, by considering $\gamma=5/3$ and
Eq.~(\ref{e:shock1}), we obtain:

\begin{equation}
P_{2}\approx 0.75(1+\alpha)\rho_{1} V_{1}^{2}
\label{e:state2}
\end{equation}

To calculate the post--shock temperature, we need of the equation of state:

\begin{equation}
P = (1+\alpha)\rho R T
\label{e:state3}
\end{equation}

\noindent where $R$ is the gas constant for the considered plasma. Therefore,
by taking into account the Eqs.~(\ref{e:shock1}), (\ref{e:state2}), and
(\ref{e:state3}), we can arrange the numerical value of the post--shock
temperature as a function of the undisturbed speed:

\begin{equation}
T_2 \approx \frac{2(\gamma -1)}{(1+\alpha)R(\gamma+1)^2} V_{1}^{2}
\label{e:state4}
\end{equation}

Please note that $V_{1}$ is given by Eq.~(\ref{e:turbo1}), by considering the
mean value. We could consider also the peak value, given by the sum of the
average plus the fluctuating turbulent value.

\section{The broadening of emission lines}
There are several models to explain the broadening of emission lines by shocks
(Perry \& Dyson 1985, Collin--Souffrin et al. 1988, Fromerth \& Melia 2001).
Recently, turbulent motion has been invoked to explain part of the broadening
(Bottorf et al. 2000), even for Fe~K$\alpha$ fluorescence line (Pairev \&
Bromley 1998). Herein we apply turbulence amplification by shocks to
investigate where in the range 0 to $10^3$ km/s (Bottorff et al. 2000) turbulence
(if it exists in AGN) may lie.

As the electron temperature of the environment plasma was measured at $10^4$~K,
if we assume that protons have the same temperature of electrons,
this implies a maximum fluctuation speed of some tens of km/s (we have
considered that under normal conditions the turbulence is limited by the local
speed of sound; cf. Balbus \& Hawley 1998).

On the other hand, we can assume that the electron temperature is not
representative of the mean temperature of the gas. If we consider a shock due
-- for example -- to a wind with speed, say $500-1000$~km/s, the post--shock
temperature, with reference to Eq.~(\ref{e:state4}), is about $1.1-4.6\cdot
10^7$~K (assuming $\alpha = 0.5$ and a hydrogen plasma). This is the ion
temperature, while the electron value is scaled by a factor $m_e/m_i\approx
5.4\cdot 10^{-4}$ ($T_e \approx 0.6-2.5\cdot 10^4$~K, a value comparable to
the observations). Indeed, it is known that the basic feature of shock waves
in plasma is the slow energy exchange between ions and electrons, and the high
electron mobility. The equilibrium temperature will be reached slowly after a
time $t_{eq}$:

\begin{equation}
\frac{dT_e}{dt}=\frac{T-T_e}{t_{eq}}
\label{e:teq}
\end{equation}

\noindent because the energy exchange occurs by Coulomb scattering and there
is large difference of mass between ions and electrons. These relaxation
processes determine the thickness of the shock front (Zel'dovich \& Raizer
1967).

From the post--shock temperature, we can calculate the most probable thermal
speed, assuming a Maxwellian distribution. Indeed, even though the electrons
and ions do not exchange energy among them, they reach independently
and quickly the Maxwellian distribution (generally after one collision).
So, the most probable speed is:

\begin{equation}
V_{th}=\sqrt{\frac{2kT}{m_i}}
\label{e:th}
\end{equation}

\noindent and the value of the turbulent speed, that is limited by the local
sound speed (cf. Balbus \& Hawley 1998):

\begin{equation}
u=c_{s}=\sqrt{\frac{\gamma kT}{m_i}}
\label{e:turb}
\end{equation}

\begin{figure*}[t]
\centering
\includegraphics[scale=0.5,angle=270]{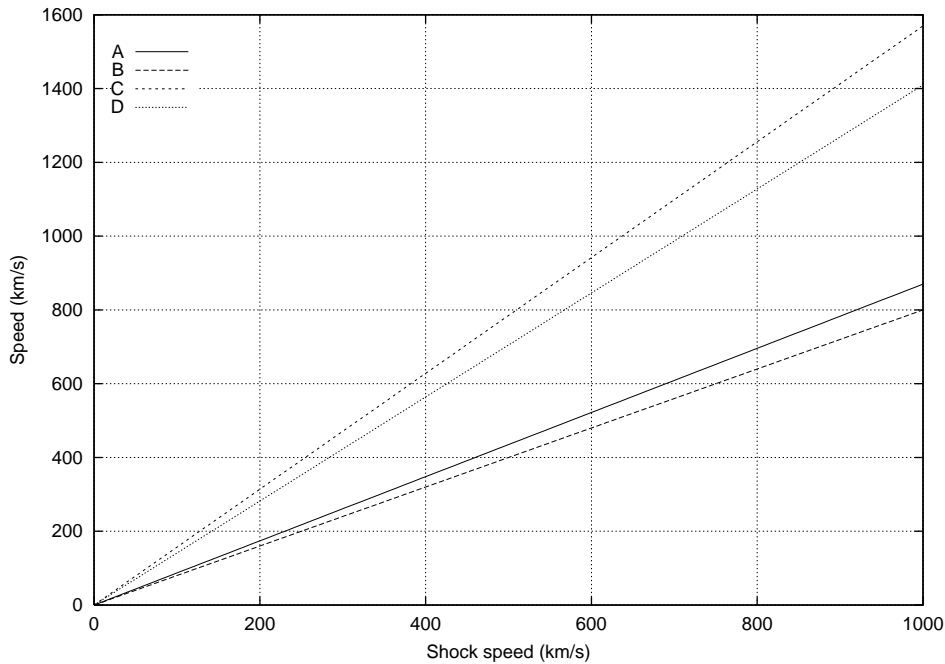}
\caption{Evaluation of post--shock thermal (A), turbulent (B), and amplified
turbulent speed (C, according to Jaquin et al. (1993); D, according to Rotman
(1991)) as a functions of the incoming shock speed.}
\label{fig:lui1}
\end{figure*}

If the shock interacts with the turbulence under unsteady conditions, we have
the amplification of the turbulent speed. We know from Sect. 3, the
amplification of the kinetic energy, from which we can calculate the amplified
speed, under the simplified assumption that the distortion affects only
the speed and not the density. We introduce the speed amplification factor $\beta$
that is simply the square root of the factor for the kinetic energy.
According to the theory by Rotman (1991), $\beta = \sqrt 2$, while for Jaquin
et al. (1993) it depends on the density ratio across the shock:

\begin{equation}
\beta = \sqrt[4]{\frac{2+\rho^2_2/\rho^2_1}{3}}
\label{e:beta}
\end{equation}

\noindent and can reach the value of $1.57$ for a monatomic gas.

Therefore, the maximum value of amplified turbulent speed is:

\begin{equation}
u_a= \beta \sqrt{\frac{\gamma kT}{m_i}}
\label{e:turb2}
\end{equation}

Substituting Eq.~(\ref{e:state4}) in Eqs.~(\ref{e:th}), (\ref{e:turb}), and (\ref{e:turb2}),
we obtain that $V_{th}\approx 0.87 V_1$, $u\approx 0.8 V_1$, and $1.41 V_1 \leq u_a \leq 1.57 V_1$,
according to the theory by Rotman or Jaquin et al., respectively. Some order--of--magnitude
calculations are shown in Fig.~\ref{fig:lui1}.

\begin{figure*}[t]
\centering
\includegraphics[scale=0.5,angle=270]{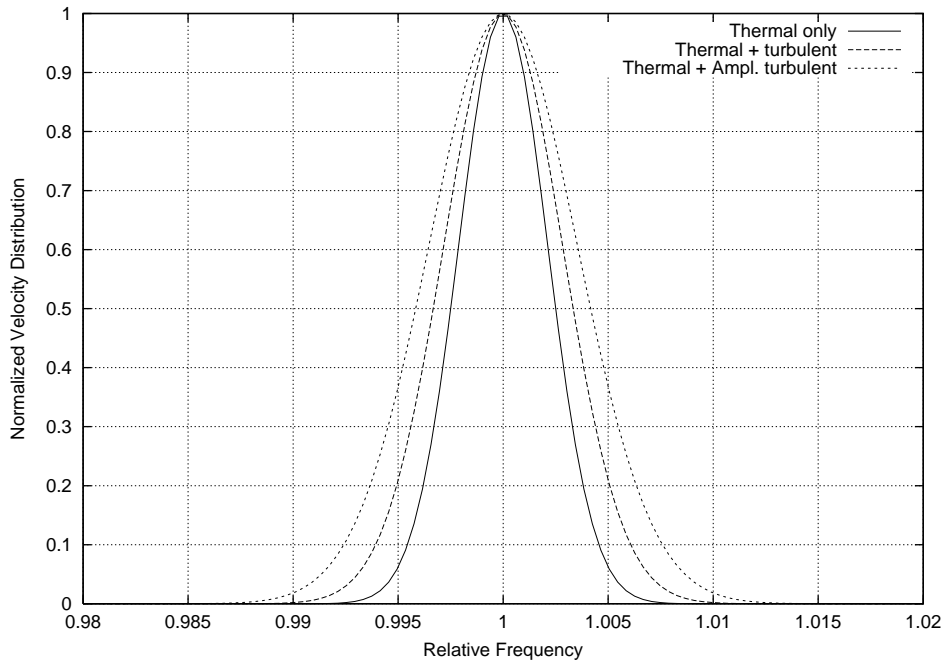}
\caption{Evaluation of broadening due to thermal speed only, additioned with
turbulence, and amplified turbulence (according to Jaquin et al. 1993, i.e. the upper limit). Values
are calculated for an incoming speed of 1000 km/s.}
\label{fig:lui2}
\end{figure*}

The maximum value of the Doppler broadening $\Delta \nu/\nu_0$ of the
emission lines due to the combined action of thermal and amplified turbulent
speed is (see Rybicki \& Lightman 1979):

\begin{equation}
\frac{\Delta \nu}{\nu_0}=\frac{1}{c}\sqrt{\frac{kT}{m_i}(2+\beta^2\gamma)}
\label{e:doppler}
\end{equation}

\noindent where $c$ is the speed of light. An evaluation of the
influence of the amplification of turbulence on the Doppler broadening
is shown in Fig.~\ref{fig:lui2} for hydrogen plasma.

We can also do the reverse calculation, i.e. starting from post--shock temperature
we can infer the speed. For example, we know that bright emission lines
stop forming efficiently when $T\approx 10^{5}$ K. For a fully ionized ($\alpha = 1$)
hydrogen plasma, Eq.~(\ref{e:state4}) gives the pre--shock speed, which has the values of about
$94$~km/s. For the amplified turbulent speed, we obtain a value between $133$ and $148$ km/s,
according to Rotman or Jaquin's theory, respectively. These values can be lower for lower
ionization: if $\alpha = 0.5$, we have that the mean value of the speed is between $115$
and $128$ km/s. Note that values change if we deal with heavy ions.

Several authors have computed models for spectra from shocks (e.g. Cox 1972, Daltabuit
et al. 1978, Shull \& McKee 1979, Binette et al. 1985) under the assumptions
of steady state and constant pressure in the post--shock region. These conditions
do not hold anymore in unsteady state and it would be necessary to set up a numerical
model to evaluate the impact of the amplification of the turbulence on the emitted
spectrum. However, we can now make some inferences: radiation generated
into the shock acts to smooth the amplification, so as to establish again a sort of
equilibrium.
We know that hydrogen line intensities are very nearly proportional to $\rho v^3$
(Cox 1972). In this note, we assumed, for sake of simplicity, that turbulence affects
only the speed and not the density, so we expect an increase of the hydrogen line
intensities due to the amplification of the turbulent speed. Moreover, because the
post--shock temperature can be high enough ($\approx 10^7$ K), we expect also an increase
of X--rays emission by thermal bremsstrahlung (if the shocked plasma is thin; otherwise
a blackbody temperature is detected).

It is worth noting, that these are order--of--magnitude calculations made with
the speed modulus, but we can observe and measure the radial component of
these vectors. So we would introduce in calculations also the line--of--sight
angle.

\section{Final remarks}
We emphasize that the purpose of this research note is to settle basic
concepts about the turbulence amplification in AGN and some of the
consequences, mainly related to the broadening of emission lines. Although, in
this case it is an additional term only, the impact of this physical
mechanism on the observations is still to be determined, particularly with
reference to LINER nuclei (cf. for example, Dopita et al. 1997, Allen et al.
1999, Evans et al. 1999).

The theory suffers from the lack of understanding of how amplification occurs
-- as underlined by several other authors -- and therefore, for this
application, observations of some AGN in soft X--rays or EUV are required to
better assess the problem.

\begin{acknowledgements}
During early stage of this research, I have appreciated helpful suggestions
in bibliographical search by J.~Frank and useful discussions with N.
Masetti. In addition, I wish to thank M. Cappi and M. Dadina for stimulating
discussions, and J.B. Stephen for a critical review of the manuscript. I thank
the four referees for their useful comments, which have helped me to improve this note.
Last, but not least, I would like to thank C. Bertout for his patience in
the editorial handling.
This research has made use of \emph{NASA's Astrophysics Data System Abstract Service}.
\end{acknowledgements}

\end{document}